\documentstyle[11pt]{article} 
\begin{document} 
\rightline{gr-qc/9708049} 
\vskip .3cm 
\vskip .3cm
\centerline{\LARGE Quantum field theory}
\vskip .2cm 
\centerline{\LARGE with and without conical singularities:}
\vskip .2cm 
\centerline{\LARGE Black holes with cosmological constant}
\vskip .2cm
\centerline{\LARGE and the multihorizon scenario}
\vskip .3cm 
\rm 
\vskip.3cm 
\centerline{Feng-Li Lin${}^a$} 
\vskip .2cm 
\centerline{Department of Physics,} 
\centerline{University of Utah,} 
\centerline{Salt Lake City, UT84112-0830, U.S.A.} 
\vskip.2cm 
\centerline{and} 
\vskip 0.2cm 
\centerline{Chopin Soo${}^b$} 
\vskip .2cm 
\centerline{National Center for Theoretical Sciences,}
\centerline{P. O. Box 2-131, Hsinchu, Taiwan 300, Taiwan.} 
\vskip .2cm  
\centerline{PACS number(s): 4.70.Dy, 4.62.+v, 4.70.-s, 4.20.Gz} 
\vskip .5cm
\centerline{Published in Class. Quantum Grav. {\bf 16}, 551-562 (1999)}
\noindent 
\vskip .2cm 
\vskip .2cm 
\vskip .2cm 
\vskip .2cm 

Boundary conditions and the corresponding states of a quantum 
field theory depend on how the horizons are taken into account. 
There is ambiguity as to which method is appropriate because 
different ways of incorporating the horizons lead to different 
results. We propose that a natural way of including the horizons 
is to first consider the Kruskal extension and then 
define the quantum field theory on the Euclidean section. 
Boundary conditions emerge naturally as consistency conditions 
of the Kruskal extension. We carry out the proposal for the 
explicit case of the Schwarzschild-de Sitter manifold with
two horizons. The required period $\beta$ is the interesting condition
that it is the lowest common multiple of $2\pi$ divided by 
the surface gravity of both horizons. Restricting the ratio 
of the surface gravity of the horizons to rational numbers yields 
finite $\beta$.
The example also highlights some of the difficulties of the off-shell 
approach with conical singularities in the multihorizon scenario; 
and serves to illustrate the much richer interplay that can occur 
among horizons, quantum field theory and topology when the cosmological 
constant is not neglected in black hole processes.

\vfill 
Electronic addresses:${}^a$ linfl@mail.physics.utah.edu; 
${}^b$ cpsoo@phys.nthu.edu.tw \hfil
\eject 
 
\section*{I. Introduction.} 

The problem of how 
quantum field theory with Schwarzschild-de Sitter (S-dS) base 
manifold\cite{de Sitter} is defined is interesting from many different angles.
Recent high redshift Type Ia supernovae observations strongly support 
the presence of a positive cosmological constant \cite{Perlmutter}.  
In black hole processes, it is physically relevant 
to take into account the effects of the cosmological constant, $\lambda$. 
It inevitably arises as the coefficient of a counterterm for quantized matter 
fields in background spacetimes. 
There are various indications that
the inclusion of the cosmological constant may affect even 
the qualitative features of black hole processes. 
Topologically, the Euclidean S-dS
manifold with two conical singularities has Euler number $\chi= 4$  and 
is not deformable to the pure Schwarzschild solution, which has $\chi =2$, 
by tuning the cosmological constant.
Naive thermodynamic arguments suggest that the pure black hole 
configuration cannot be obtained as the smooth thermodynamic 
limit $\lambda \rightarrow 0$ 
of the S-dS configuration with two horizons since 
the size of the outer cosmological horizon becomes infinitely large then 
and should contribute infinite entropy as the cosmological constant goes 
to zero, whereas the pure black hole lacks the outer horizon altogether. 

In recent years, methods have been developed to regularize 
the contributions of conical manifolds \cite{Dowker,Miele}.   
They allow, for instance, the discussion of the thermodynamics of black 
holes in the off-shell approach in which the Hawking temperature 
for the Schwarzschild black hole is derived from the 
thermal equilibrium condition given by the extremum of the Euclidean 
Einstein-Hilbert action \cite{Fursaev}. 
The off-shell approach also makes it feasible to decouple
the inverse of the temperature, $\beta$, from the Hamiltonian which depends 
on the mass of the black hole by lifting the on-shell 
restriction $\beta = 8\pi m$.
If conical singularities are allowed, we can consider more complicated
scenarios to test if the formalism leads to difficulties which are not 
encountered in the case of the pure black hole. S-dS with two 
horizons appears to be a particularly relevant example.

In the region between but not including the horizons of S-dS, a single global 
coordinate patch exists. {\it The question is how one takes into account the 
horizons.}
In the method with conical singularities, the horizons of S-dS are to be
included as conical singularities. However, an important difference for this 
multihorizon situation is that there is no straightforward way to define 
the usual on-shell thermal equilibrium temperature as in the case of 
the pure black hole\cite{Hawking} because it is 
impossible to simultaneously eliminate 
the conical defects on both horizons by a single choice of periodicity 
within the formalism with conical singularities. From this perspective,
the strategy of allowing for conical singularities 
therefore seems rather pertinent and also needed for a multihorizon 
scenario such as the S-dS since it 
permits adopting a {\it single} off-shell periodicity which does not need to 
coincide with either of the values required to remove conical defects 
at the horizons. It also seems to imply that an off-shell discussion
of thermodynamics is possible despite the apparent unequal intrinsic 
periodicities of the horizons.

However, as we shall see, this is not the only way to resolve the impasse.
Actually, the method does not seem to give the correct results even for the 
pure de Sitter case. There are also questions with regard to 
the consistency, or at least ambiguity, of quantum field theories 
defined on manifolds with conical singularities. 
In the one-loop efffective action in background
spacetimes, there are terms involving the square of the curvatures 
which are divergent and cannot be removed in the off-shell approach with 
conical singularities 
precisely because in this formulation, no single choice of $\beta$ can 
simultaneously get rid of both conical defects at the horizons. 
For the pure black hole, one can take
the on-shell limit after the computations are done to eliminate 
the unwanted terms\cite{Fursaev}.

In Ref.\cite{Hawking}, it is suggested that we can consider partitioning
 the volume into two regions which are in equilibrium with the 
respective inner and outer horizons. 
We are then able to do thermodynamics without 
conical singularities. However, the partition is by no means natural. 
Moreover, it is very much unlike a 
patching condition in that the physics depends on how the partition 
is chosen. Half of the total volume at each natural temperature of 
the horizons is clearly different from one-third of the volume at 
one temperature and two-thirds of it at the other. 
On the other hand, we may even argue that the physical situation of S-dS may 
correpond more closely to a situation with temperature gradient and even
non-equilibrium physics since the natural 
surface temperatures of the horizons are different. 

It is interesting to note that either extremes can have
dramatic implications for black hole processes.
If conical singularities are allowed, they may be potentially 
significant, both as remnants of black hole evaporation 
and seeds for black hole condensation in a de Sitter universe
with conical singularities, and can 
actually serve to preserve the information of the topological Euler number 
during these processes. It may be possible 
for a black hole of the S-dS type to achieve the zero 
mass limit with two conical singularities and a remaining outer horizon, 
and still maintain the $\chi = 4$ condition. Moreover the remaining 
outer horizon could be larger than the sum of the initial black hole and 
cosmological horizons. This could be consistent with information loss without
violating topological conservation laws. On the other hand,
if conical singularities are to be excluded, then the mere introduction of 
the cosmological constant, which can also be induced 
from quantized matter, could lead to non-equilibrium processes with 
deviations from blackbody spectrum and its implications for 
the information loss paradox, due 
to the presence of two horizons with unequal surface gravity.
But neither of these simple extreme scenarios may be entirely correct.
The issue of how to define, say quantum field theory, on such a background 
with multiple horizons has yet to be settled.

The imposed boundary conditions and corresponding states 
of the quantum field theory depend on how the horizons are accounted for. 
So it is pertinent to ask if there are {\it natural} ways to incorporate 
the horizons. 
In this paper, we compare the scenarios with and without conical
singularities and illustrate some of the difficulties that are present
in the former. 
We return to the Kruskal extension of the pure black hole solution and 
observe that there is a generalization for S-dS which will {\it naturally} 
incorporate the horizons. The Euclidean quantum field theory is then 
defined without conical singularities but with patching and consistency 
conditions which determine the feasible states. When applied to the
pure black hole and $S^4$ de Sitter configurations, the proposal
yields the correct Gibbons-Hawking temperatures.

\section*{II. Conical singularities and QFT in S-dS spacetime.} 

In finite temperature quantum field theories in flat spacetime, 
temperature dependence of the effective action is introduced through 
radiative loop corrections and resummation \cite{Kapusta}. However, 
in curved spacetimes there is temperature dependence in the action even 
at tree level through the periodicity of the Euclideanized metric. 

In the formulation with conical singularities, {\it the horizons are 
accounted for as conical singularities}\cite{Fursaev}.
The Euclidean Einstein-Hilbert action which includes contributions from
conical singularities is
\begin{equation} 
I_g =-{1\over 16\pi}\int_{M/\Sigma} d^4 x {\sqrt g}(R-2\lambda)  
-{1\over 16\pi}\int_{\Sigma} d^4x {\sqrt g}R -{1\over 8\pi}\int_{\partial M} 
d^3x {\sqrt h}(K-K_0). 
\end{equation} 
Here, $\Sigma$ denotes the singular set of the horizons due to conical 
defects, and $K$ is the second fundamental form.
 
The partition function and effective action are defined through 
the path-integral with
\begin{equation} 
e^{-I_{eff}(\beta)}=Z(\beta)=\int [{\cal D}g][{\cal D}\phi]e^{-I_g-I_m} . 
\end{equation} 
It is assumed that the period of the Euclidean time variable 
(which does not need to coincide with either of the values required 
to remove conical defects at the horizons) is $\beta \equiv 1/T$. 
$I_m$ is the matter action for $\phi$ while $I_g$ is
the gravitational action. Thermodynamic information may be extracted from 
the partition function. For example, Fursaev et al \cite{Fursaev} derived 
the Hawking temperature and 
Bekenstein-Hawking entropy for the pure Schwarzschild black hole 
through this formulation from the extremum of the effective action.

We are interested in the case of a black hole with 
positive cosmological constant
i.e. the S-dS configuration, and the Euclidean region between the two 
horizons. There are two horizons with the larger cosmological horizon 
at $r_+$ and the inner black hole horizon at $r_-$ if we impose the 
restriction $9m^2 \lambda < 1$ \cite{de Sitter,Hawking}. 
The region between the inner black hole and outer cosmological horizons
also serves as a natural volume for thermodynamic considerations. 
For the Euclidean S-dS configuration,
there are no boundary terms in Eq.(1). 
This is in contradistinction with the pure 
Schwarzschild case where the boundary term at infinity contributes to the 
Arnowitt-Deser-Misner mass of the black hole. 
The Euclidean section of interest 
has a conical singularity at each horizon and their contributions 
to the Einstein-Hilbert action are taken into 
account by the second term in Eq.(1). 

Specifically, the Euclidean S-dS metric is 
\begin{eqnarray} 
ds_{E}^2 = h(r) d\tau^2 + {dr^2 \over h(r)} + r^2 d\Omega^2 , 
\\ 
h(r)=1- {2m\over r} - {\lambda r^2 \over 3}. 
\end{eqnarray} 
Here $\tau$ is the periodic coordinate with periodicity equal to 
$\beta$.
As stated earlier, there are conical singularities on the horizons 
when $T$ is not equal to the individual 
Gibbons-Hawking temperatures associated with 
the horizons. To reveal the conical singularities on the 
horizons, we may choose the local coordinate patches and change variables 
through
\begin{equation}
h(r) = k^2 X^2.
\end{equation}
The metric becomes 
\begin{equation}
ds^2_{E} = X^2 d\left({k\tau}\right)^2 
+ {dX^2 \over ({ h^{'} \over {2k}})^2} + r^2 d\Omega^2.
\end{equation}
$V^{'}$ denotes the first derivative of $V(r)$ with respect to $r$. 
Near the horizons at $r_{\pm}$, the topology reduces to ${\cal C}^2 \times \
{\cal S}^2$ if we set  
\begin{equation} 
k_\pm = \frac{1}{2}\left| {h^{'}(r_\pm)} \right|. 
\end{equation} 
$r_{\pm}$ are the solutions of $h(r_\pm)= 0$ and are related 
to $m$ and $\lambda$ through
\begin{equation} 
r_+ ={\sqrt \frac{4}{\lambda}}\cos(\frac{\xi +4\pi}{3}), \qquad 
r_- ={\sqrt \frac{4}{\lambda}}\cos(\frac{\xi}{3}), \qquad 
\cos(\xi)= -3m{\sqrt \lambda} , 
\end{equation} 
with $\xi$ in the range $(\pi,\frac{3\pi}{2}]$ and, 
$0 \leq 9m^2\lambda < 1$ \cite{de Sitter}. 
Note that $k_\pm$ are the values of the surface gravity\footnote{
It may be more natural to include a normalization factor in defining the
surface gravity(see for instance Ref.\cite{Bousso}). We thank R. Bousso
for drawing our attention to this. However, according to Eqs.(A6)-(A9) 
of Ref.\cite{Bousso} the factor cancels in required periods and therefore 
none of our conclusions will be affected.} on the 
horizons, and the Gibbons-Hawking temperatures $T_\pm$ associated with the 
respective horizons are  
\begin{equation} 
T_\pm={k_\pm \over 2\pi}\label{local temp}. 
\end{equation} 
It is thus clear for the manifold defined this way that there are conical 
defects if $T$ is not equal to $T_\pm$. 

Following Ref.\cite{Fursaev}, the contributions of the conical 
singularities to the action are 
\begin{eqnarray} 
-{1\over 16\pi}\int_{\Sigma} d^4x {\sqrt g}R &=& 
-{1\over 4}(1-{T_- \over T})A_- 
+{1\over 4} (1-{T_+ \over T})A_+. 
\end{eqnarray} 
The relative sign difference in the above 
expression reflects the opposite orientations of the normals 
at the horizons with respect to $dr$. 
The conical contributions vanish when the area of the corresponding 
horizons $A_{\pm}$ coincide and $T_- = T_+$. 
The conical contributions are obviously nontrivial otherwise.

We also need the contribution from the non-singular set $M/\Sigma$ 
to compute the total contribution to $I_g$ in Eq.(1). 
By integrating $r$ from $r_-$ to $r_+$ and $\tau$ from $0$ to $\beta$, 
the contribution to the Einstein-Hilbert action from the non-singular 
set is
\begin{eqnarray}
-{1\over {16\pi}} \int_{M/{\Sigma}} d^4x \sqrt{g}(R-2\lambda)& =&
-{\lambda \over {8\pi}} \int_{M/\Sigma} d^4 x \sqrt{g}\cr
\nonumber\\
&=&-{{\lambda\beta} \over 6}(r^3_+- r^3_-).
\end{eqnarray}
Summing the contributions of Eqs.(10) and (11), we have the tree level 
action (with $I_m \equiv 0$) with temperature dependence as  
\begin{equation} 
I_g=-\beta(\lambda r_+^3/3 - m)-\pi (r_-^2 - r_+^2).  
\end{equation} 
We may consider extremizing and doing thermodynamics with this action,
but it does not seem to yield sensible results in the off-shell approach.
For instance, the entropy in this approach with conical singularities
is 
\begin{equation}
S_{Con. sing.}= \beta {\partial I_g \over {\partial \beta}}- I_g 
= \pi (r^2_- - r^2_+).
\end{equation}

At this point, it is appropriate to spell out a few subtleties and 
difficulties associated with the formulation with conical singularities. 
First of all, there is a subtlety with regard to the 
correct sign of the action.
When the entropy is evaluated using this approach for 
the pure de Sitter configuration\footnote{The $S^4$ de Sitter 
configuration is interesting from the thermodynamic viewpoint
in a number of ways.
In quantum gravity, it may be necessary to have a nonvanishing 
cosmological constant. The Gibbons-Hawking temperature of the de Sitter 
solution, which is the configuration with the greatest symmetry and a 
possible ground state in quantum cosmology, is proportional to 
$\sqrt \lambda$, and the exact vanishing 
$\lambda$ limit may be a physically unattainable zero temperature limit in 
quantum gravity \cite{smolin}. 
The de Sitter solution also 
appears to violate Nernst's theorem explicitly 
since its entropy which is proportional to the area of the horizon and 
inversely proportional to $\lambda$ does not go to zero with
vanishing temperature.}, the result is {\it negative}. Explicitly,
\begin{equation}
S_{Con. sing.}= \beta {\partial I_g \over {\partial \beta}}- I_g 
= -\pi r^2_+.
\end{equation}
The absolute value is the correct Gibbons-Hawking
entropy for pure de Sitter manifold with cosmological horizon at $r_+$.
This is in contradistinction with the pure Schwarzschild case where the 
action as in Eq.(1) gives the correct sign and magnitude 
of the Bekenstein-Hawking entropy and also the correct positive 
energy equal to the Arnowitt-Deser-Misner mass $m$\cite{Fursaev}.
We emphasize that {\it on-shell} calculations for the de Sitter 
solution with $\beta = 2\pi r_+$
gives the correct positive result for the entropy 
because $I_g$ in Eq. (1) {\it without conical singularities} leads to
$I_g = -S = -\pi r^2_+$\cite{Euclidean}.
We may try to choose the action to be
the negative of that in Eq.(1)
but that convention will lead to problems with the pure Schwarzschild case. 
The entropy calculated from the method with conical singularities 
therefore may or may not coincide with the on-shell value even in 
situations where there is but a single horizon. Moreover, it can even lead 
to non-positive values of $S_{Con. Sing.}$.
 
Secondly, there are difficulties associated with the formulation with
conical singularities if we were to apply it to QFT of matter fields in 
curved spacetimes with more than one horizon. Physically, it is important to 
include matter but from Eq.(2), on integrating out the quantum field 
$\phi$ in a fixed background metric, the effective action is naively 
expected to be \cite{Birrell}  
\begin{eqnarray} 
I_{eff}[g_{\mu\nu}, \beta,\lambda]&=&
\int_M d^4x {\sqrt g}\{{-1 \over 16\pi G_{ren}}
(R- 2\lambda_{ren})\cr
\nonumber\\ 
&+&c_1 R^2+c_2 R^{\mu \nu} R_{\mu \nu}+c_3 R^{\mu \nu \alpha \beta} 
R_{\mu \nu \alpha \beta}\} + {\rm finite \,\, terms}.
\end{eqnarray} 
This is for smooth manifolds.
Curvature-squared quantum corrections also contribute to the conformal anomaly
which is related to the Hawking radiation, and are thus physically 
relevant to the thermal feature of spacetime \cite{Christensen}. 
However, when 
conical singularities are present, they result in Dirac $\delta$ 
singularities in the curvatures\cite{Fursaev}. Components
of the curvature tensor have to be defined as distributions and
integral characteristics of quadratic and higher powers of the curvature
do not have strict meaning. We may assume all the higher-order 
renormalized coefficients of curvature-squared terms vanish identically to 
bypass this difficulty i.e. to assume that the only effect of quantized matter 
is just to renormalize the gravitational constant $G$ (which has been set 
to unity in our convention) and the cosmological constant $\lambda$. 
However this is due more to expediency than to 
compelling physical arguments. 
As was pointed out in Ref.\cite{Miele}, the trace of the heat kernel operator
turns out to be well-defined, and we may compute the QFT contributions
from the asymptotic expansion of the trace of the heat kernel operator
\begin{equation}
Tr(K) = Tr(\exp(-s\triangle)) = {1\over {4\pi s}^2}
(a_0 + s a_1 + s^2 a_2 +...).
\end{equation}
However it is still necessary to {\it assume} potential terms in 
the Laplacian operator are defined only on $M/\Sigma$ and do 
not include singular terms. In general, the conical contributions
to the coefficients $a_i$ do not vanish. 
Moreover, for spin 3/2 and spin 2 fields, even when the on-shell
value of $\beta$ is taken afterwards, the trace of the heat kernel 
differs from the trace on smooth manifolds.
In the case of the pure black hole where there is but a single horizon, 
``renormalizations"\footnote{
In the case of quantized matter in Schwarzschild {\it background}
of fixed mass, the black hole mass is treated as a macroscopic 
parameter so that $Z(\beta, G, m)$.} can be done and the contributions of 
curvature-squared terms at equilibrium temperature (at which the conical 
defect disappears) can be taken into account \cite{Fursaev}.
The crucial difference is that this cannot be done for the relevant 
multihorizon scenario here because it is impossible to {\it simultaneously} 
eliminate conical defects at both horizons.

While these hurdles do not conclusively show that there is no 
sensible way to define QFT in S-dS spactime via the conical method, it is
nevertheless true that that different ways of accounting for the horizons
 lead to different results. There is thus ambiguity as to which
of the methods is ``appropriate". Some of the methods are covered in
a review of on-shell vs. off-shell computations\cite{Offshell}. The discussion 
includes the ``brick-wall" method with Dirichlet boundary conditions and 
a cut-off distance from the horizon, the ``blunt cone" method where the
conical singularities are smoothened away by a deformation parameter, 
the ``volume cut-off" formalism, the method with conical singularities, 
and on-shell computations for the pure black hole.

Actually, the very meaning of ``on-shell" for the 
case of Schwarzschild-de Sitter with two horizons is 
problematic from the perspective of the conical method, and is 
so far undefined. But this will be pursued in the next section.

\section*{III. Kruskal extension of the S-dS spacetime.} 

We emphasize that different ways of accounting for the horizons
lead to different results for the effective action.
We propose that a more natural way to account for the horizons 
is to consider the Kruskal extension of the manifold and 
then define QFT on the Euclidean section. 
We draw a lesson first from 
Kruskal extension of the pure Schwarzschild solution and see why
it leads to $\beta = 8\pi m$ naturally.

(a) The pure black hole metric is
\begin{equation}
ds^2 = -h(r)dt^2 + h^{-1}(r) dr^2 + r^2 d\Omega^2
\end{equation}
with $h(r) = {{(r - 2m)} \over r}$.
By defining $u = t- r^* $ and $v= t+ r^*$, with
\begin{eqnarray}
r^* &=& \int {r\over (r-2m)} dr \cr
\nonumber\\
&=& r + 2m\ln (r-2m),
\end{eqnarray}
the metric can be transformed to  
\begin{equation}
ds^2 = -h(r) du dv + r^2 d\Omega^2.
\end{equation}
The Kruskal extension can be done with coordinates
\begin{equation}
u' =  -e^{-ku}, \qquad v' = e^{kv}
\end{equation}
where $k = {1\over 2} {dh/dr}|_{r=2m} = \frac{1}{4m}$ is the surface gravity 
at the  horizon.
In terms of Kruskal coordinates, the metric becomes
\begin{eqnarray}
ds^2&=& -h(r){du \over du'}{ dv \over dv'} du' dv' + r^2 d\Omega^2 \cr
\nonumber\\
&=& -{1\over k^2 r}e^{-2kr} du' dv' + r^2 d\Omega^2.
\end{eqnarray}
By rewriting
\begin{equation}
t' = (u' + v')/ 2 \qquad r' = (v' - u') /2, 
\end{equation}
we have
\begin{equation}
ds^2 ={1\over {k^2r}}e^{-2kr}({dr'}^2 -{dt'}^2)  + r^2 d\Omega^2
\end{equation} 
and $ {r'}^2 - {t'}^2 = -u' v' =  e^{2kr}(r - 2m) $. 
So the Euclidean section for which the metric is positive-definite 
can be defined by a Wick rotation of $t$ which makes 
$t'$ pure imaginary; and hence for $r \geq 2m$ only. 
Moreover, $\tau = it $ has period $2\pi/ k$ since
\begin{equation}
 t'= (u' + v')/2=  e^{k r^*}\sinh(-ik\tau),
\end{equation}
and $(u', v')$ defines $\tau$ up to multiples of $2\pi/ k = 8\pi m$.
This periodicty for $\tau$ is precisely the condition for the 
manifold to be free of conical singularity, 
but {\it it emerges naturally as a consistency condition of the 
Euclidean section of the Kruskal extension.} In the Kruskal extension,
the form of the metric is non-singular at the horizon. Euclidean QFT can be
constructed for the Kruskal extension with the on-shell restriction
of $\beta = 8\pi m$.

Note that if we neglect the spherically symmetric $d\Omega^2$-part, only
one coordinate patch is required for the Kruskal extension because the
topology is $R^2$ and the topology of the four-manifold is $R^2 \times S^2$.
The Euclidean section has Euler number $\chi =2$.
For the Schwarzschild-de Sitter metric, the Kruskal extension needs more 
than one coordinate patch even if we neglect the spherically symmetric part
of the metric because there are now two horizons with unequal surface
gravity. However it is clear that the Euclidean section of the Kruskal
extension of the black hole {\it includes the horizon} and also 
yields {\it consistency conditions} on the periodicity of $\tau$. 
We shall therefore use the Kruskal extension to consistently incorporate
both horizons of the Schwarzschild-de Sitter manifold in the Euclideanization,
and also to deduce the consistency conditions that are required.
In this manner, boundary conditions for quantized matter fields in the
Schwarzschild-de Sitter background will emerge naturally from the Euclidean
quantum field theory. Moreover, as we shall see, the conditions that 
arise are rather interesting.

(b)The Schwarzschild-de Sitter metric has the form
\begin{equation}
ds^2 = -h(r)dt^2 + h^{-1}(r)dr^2 + r^2 (d\theta^2 + \sin^2\theta d\phi^2 ),
\end{equation}
with
\begin{equation}
h(r) = {\lambda \over {3r}}(r_+ - r) (r- r_-)(r+ r_+ + r_-).
\end{equation}
It is a solution of Einstein's equations with cosmological constant 
$\lambda$ if
\begin{equation}
3/\lambda = r^2_+ + r_+r_- + r^2_- ,\qquad 6m/\lambda = r_+r_-(r_+ + r_-).
\end{equation}
The metric is a priori defined for the region between the horizons but there 
can be a Kruskal extension.\footnote{See, for instance, Ref.\cite{Tadagi}.}

The surface gravity at the horizons are given by
\begin{equation}
k_{\pm} = {1\over2 }\left|dh(r)/dr\right|_{r = r_\pm}, 
\end{equation}
or
\begin{equation}
k_{\pm} = {\lambda\over {6 r_{\pm}}} (r_+ - r_-)(2r_{\pm} + r_{\mp}).
\end{equation}
The horizons have different values of surface gravity and their  
ratio satisfy $ 0 < k_+/k_- \equiv \alpha \leq 1$.

Similarly, we define
\begin{equation}
u \equiv t - r^* \qquad v \equiv t + r^*
\end{equation}
with
\begin{eqnarray}
r^*&=& \int h^{-1}(r) dr \cr
\nonumber\\
&=& {1\over {2k_-}}\ln(r- r_-) -{1\over {2k_+}}\ln(r_+ - r) 
+ ({1\over {2k_+}} - {1\over {2k_-}})\ln (r + r_+ + r_-).
\nonumber\\
\end{eqnarray}
In terms of these coordinates,
\begin{equation}
ds^2 = -h(r(u, v))dudv + r^2(u, v) (d\theta^2 + \sin^2\theta d\phi^2 ),
\end{equation}
with $r$ defined implicitly by $ r^*(r) = (v-u)/2$.

We may cover the Kruskal extension by two coordinates patches 
$(u_{\pm}, v_{\pm})$.
$(u_+, v_+)$ is valid for $ r > r_- $, which includes the outer or
cosmological  horizon but not the inner or blackhole horizon;
while $(u_-, v_-)$ is valid for $ r < r_+ $, and includes the inner but
not the outer horizon.
These coordinates are 
\begin{equation}
u_\pm = \pm e^{\pm k_{\pm} u}, \qquad v_\pm = \mp e^{\mp k _\pm v}.
\end{equation}
Thus
\begin{eqnarray}
du_{\pm} dv_{\pm} &=& k^2_\pm e^{\pm k_\pm (u-v)} du dv \cr 
\nonumber\\
&=& k^2_\pm e^{\mp 2k_\pm r^*}(dt^2 - {dr^*}^2),
\end{eqnarray}
and
\begin{eqnarray}
ds^2 &=& - h(r){du \over du_\pm }{dv \over dv_\pm}du_\pm dv_\pm 
 + r^2 d\Omega^2 \cr
\nonumber\\
&=&  -h_{\pm} du_\pm dv_\pm + r^2 d\Omega^2 ,
\end{eqnarray}
where
\begin{equation}
h_- = {\lambda\over 3k^2_- r}(r_+ - r)^{(1+\alpha^{-1})}
(r + r_+ + r_-)^{(2-\alpha^{-1})}, 
\end{equation}         
and
\begin{equation}
h_+ = {\lambda\over 3k^2_+ r}(r - r_-)^{(1+\alpha )}
(r + r_+ + r_-)^{(2-\alpha)}.
\end{equation}         
$h_-$ is valid for the patch with $0 < r < r_+$ and $h_+$ for that with 
$r > r_-$.
Therefore it is clear that the metric is nonsingular (except at $r=0$) 
and also nonvanishing in each of the respective coordinate patch.
The overlap of the patches occurs in the region $ r_- < r < r_+$ 
where the coordinates are related by 
\begin{equation}
u_+ = -e^{(k_+ + k_-)u}u_- , \qquad v_+ = -e^{-(k_+ + k_-)v} v_- .
\end{equation}

We may also note that by a Wick rotation of $t$, 
the metric becomes positive definite for $r_- \leq r \leq r_+$
since with $\tau = it$ the metric is
\begin{equation}
ds^2 = h_\pm k^2_\pm e^{\mp 2k_\pm r^*}(d\tau^2 + {dr^*}^2) 
+ r^2 d\Omega^2.
\end{equation}
In terms of Kruskal coordinates $(u_\pm, v_\pm)$, the metric is
\begin{equation}
ds^2 = h_\pm |du_\pm|^2 + r^2 d\Omega^2.
\end{equation}
since $u_\pm$ become complex conjugates of $-v_\pm$ after Euclideanization.
To satisfy $|u_\pm|^2 \geq 0$, the Euclidean section is defined only 
for $ r_- \leq r \leq r_+$; and it can be shown that the horizons at $r_\pm$ 
correspond to the origins $u_\pm = 0$. Expression (40) shows that the 
extended Kruskal Riemannian manifold exhibits no singular behaviour at 
the horizons.
In the overlap region with $r_- < r < r_+$,
\begin{equation}
\left[\matrix{u_+ \cr  v_+}\right]
= \left[\matrix{-e^{(k_+ + k_-)(-i\tau - r^*)} & 0\cr
0 &-e^{-(k_+ + k_-)(-i\tau + r^*)} }\right] \left[\matrix{u_- \cr
v_-}\right].
\end{equation}
So the transition function 
is single-valued  only if $\beta $, the period of $\tau$, is an integer
multiple of ${2\pi} \over {(k_+ + k_-)}$. 
However, there are stronger consistency conditions. Since
\begin{equation}
(u_\pm + v _\pm)/2 =  e^{\mp k_\pm r^*} \sinh(-ik_\pm \tau),
\end{equation}  
this means that
$(u_\pm, v_\pm)$ only define values of $\tau$ up to integer multiples 
of $2\pi /k_+$ and $2\pi/k_-$ in each patch. But 
$(u_\pm, v_\pm)$ are also well-defined coordinates in the overlap.
Therefore the translation $\tau \rightarrow$ $\tau + \beta $ which leaves 
{\it both} sets of coordinates $(u_\pm, v_\pm)$ invariant must be such that
$\beta$  has to be an integer multiple of {\it both} $2\pi/k_+ $ and
$2\pi/k_-$ .
This means that in fact  $\beta$, the period of $\tau$, 
is therefore the {\it lowest common multiple} of ${2\pi} \over {k_+} $ 
and ${2\pi} \over {k_-}$.
It is easy to check that this is sufficient (although not necessary) for 
the transition function to be single-valued. The latter is a weaker 
condition. There is however an interesting relation: Let $n_\pm$ be 
relatively prime positive integers 
such that $\beta \equiv  2\pi n_\pm /k_\pm$. Thus
 $ 0 < \alpha = k_+/k_- = n_+/n_- \leq 1$ is rational.\footnote{The 
metric of Eqs.(35)-(37) is still well-defined for irrational exponents through 
$a^x = \exp(x\ln a) ,  a > 0$. Restricting $\alpha = k_+/k_-$ to rational 
numbers yields finite $\beta$. Rational numbers are also dense in the system
of real numbers. If $\alpha$ is irrational, $\beta$ becomes larger and larger 
with improving approximations of irrationals by rationals.}
Then $(n_+ + n_-) = {\beta\over {2\pi}}(k_+ + k_-)$ or
\begin{equation}
\beta = 2\pi {{(n_+ + n_-)}\over{(k_+ + k_-)}}. 
\end{equation} 
By comparing with
Eqs.(38) and (40), we see that under $\tau \rightarrow \tau + \beta$, 
the transition functions of the $u_\pm$ and $v_\pm$ coordinates 
 gets multiplied by $\exp[\mp i2\pi(n_+ + n_-)]$; and 
$(n_+ + n_-)$ is the {\it winding number} of the transition function.

\section*{IV. Topological considerations.}

The Lorentzian Kruskal extension of S-dS is known to exhibit a multi-sheeted 
structure with the Penrose diagram showing repeating units \cite{Euclidean}. 
Thus there is the question of what one means by the Euclidean section with 
$r_- \leq r \leq r_+$, 
and what the actual topology (specifically
the Euler number) is.
The consistency condition that we uncovered in the previous section 
is related to these issues.

We first compute the Euler 
number of the Euclidean manifold 
{\it with conical singularities} \cite{Fursaev}.
This is given by 
\begin{equation}
\chi[M]=\chi[M/\Sigma]+\chi[\Sigma],
\end{equation}
with the regular contribution 
\begin{equation}
\chi[M/\Sigma]= {1 \over 32\pi^2}\int_{M/\Sigma} d^4x {\sqrt g}
(R^2-4R_{\mu \nu}^2+R_{\mu \nu \alpha \beta}^2),
\end{equation}
and the contributions from the conical singularities,
\begin{equation}
\chi[\Sigma]=\sum_\pm (1-\frac{\beta}{\beta_\pm})\chi_2[\Sigma_\pm].  
\end{equation}
$\beta_\pm = 1/T_\pm = 2\pi/ k_\pm$ are the periods associated 
with the horizons.
For the Euclidean S-dS manifold, the explicit computation is 
straightforward. The results are
\begin{eqnarray}
\chi[M/\Sigma]&=& 2\beta({1\over \beta_+} + {1\over \beta_-}),
\\
\chi[\Sigma]&=& \chi_2[{S}^2]
[2-\beta({1\over \beta_+} + {1\over \beta_-})]
\nonumber\\
&=&  4 -2\beta({1\over \beta_+} + {1\over \beta_-})
\end{eqnarray}
since $\chi_2[{S}^2] = 2$.
Thus, as expected, $\chi[M]$ for S-dS is {\it always} equal to $4$ 
and is {\it independent of $\beta$} if the horizons are incorporated as 
conical singularities\footnote{For the pure Schwarzschild black hole, similar 
computations yield $\chi=2$.}. In this sense, allowing for conical
singularities preserves the toplogical information which
remains constant under deformations of the parameters $m , \lambda$ 
and $\beta$; and conical singularities seem rather appealing 
as seeds for condensation and remnants of black hole evaporation.
For instance, the specific relation between $\beta$ and the
value of $m$ which extremizes (at fixed $\lambda$) the action of Eq.(12) 
can be worked out. There are actually critical values of $\beta$ for 
which $m$ approaches $0^+$ and becomes larger as $\beta$ is varied. 
However, when we wish to consider higher order curvature terms,
there are ambiguities and difficulties associated with QFT contributions 
if horizons are accounted for as conical singularities.

In contrast, the Euclidean section of the Kruskal extension gives a
different result for the Euler number. Recall that $\beta$ has to be 
the lowest common multiple of $2\pi/k_+$ and $2\pi/k_-$ to satisfy the
consistency condition discussed in the previous section so that
the translation $\tau \rightarrow \tau + \beta$ is a symmetry.
In the Kruskal extension there are no conical singularities, and
Eq.(44) yields
\begin{equation}
\chi[M]= 2\beta({1\over \beta_+} + {1\over \beta_-})
= 2(n_+ + n_-).
\end{equation}
In the last step above, we have substituted the values of $\beta$ from 
Eq.(43) and $\beta_\pm = 2\pi/k_\pm$.
The Euler number is therefore integer and even. 
The former is consistent with the Euler number of Riemannian manifolds, 
and the latter is due to the spherical symmetry as $\chi_2[{S}^2] =2$. 
As mentioned previously, $(n_+ + n_-)$ is the winding number of the
transition function displayed in Eq.(41). Therefore we may also write
Eq.(43) as
\begin{equation}
\beta = {{\pi \chi}\over{(k_+ + k_-)}}.
\end{equation}
However the Euler number is not fixed to be exactly 4.
 The reason is that basic repeating Euclidean 
unit for which $\tau \rightarrow \tau +\beta $ is a 
symmetry depends on both $k_+$ and $k_-$.

In the approach with conical singularities, the Euler number is
divided between the singular and regular parts of the manifold.
These two values can be adjusted by changing $\beta$ although
their sum is always 4.
Computations of other invariants further differentiate between 
the alternatives. With conical singularities, the four-volume is
${{4\pi \beta} \over 3} (r^3_+ - r^3_-)$ and is a function of $\beta$
which is independent,
while $\beta$ is not arbitrary for the case with Kruskal extension.
In the method with conical singularities, conical contributions to 
the action given by Eq.(10) also do not vanish in general.
Thus even if some invariants can be matched by certain choices of
$\beta$, others will not be.
Only the limiting case of $k_+ = k_-$ allows for a correspondence 
between the formalism with conical singularities and the
results from using the Kruskal extension, 
since in this limiting case conical defects on both horizons can be 
eliminated by a single choice of $\beta$ in the formalism with 
conical singularities.

\section*{V. Remarks}

We have discussed some of the difficulties with 
QFT contributions in the off-shell approach 
if the horizons are to be accounted for as 
conical singularities of the Euclidean section.
The problems becomes more transparent and acute in scenarios with more 
than one horizon; and we have considered the explicit example 
of Schwarzschild-de Sitter.  
A more natural way to incorporate the horizons emerges from considering 
the Kruskal extension and then constructing the QFT on the Euclidean 
section.
In this manner no conical singularites are introduced but the horizons 
with their unequal surface gravity lead to natural selection rules or
consistency conditions on the periodicity of the Euclidean time variable;
and suggests that these are the natural boundary conditions that
should be imposed upon such a QFT and the quantum states. 
Moreover, this implies that thermal states with  $\beta$ being lowest 
common multiple of $2\pi/k_+$ and $2\pi/k_-$ exist. 
In this approach, $\beta$ can no 
longer assume arbitrary off-shell values but is completely determined
 by the stated consistency condition.
Since there are no conical singularities, the one-loop
effective action on integrating out quantized matter fields will
contain the usual terms (and counterterms) without arbitrary
$\beta$-dependent contributions and Dirac delta singularities in
terms quadratic in the curvatures.
Although we have not set up an explicit quantum field theory and 
completed the calculation of the stress tensor in 4-d, 
there is support for our conjecture. The existence of quantum states 
for S-dS whose stress tensor is static has been shown explicitly for the 
2-d case of S-dS for which the angular dependence is neglected \cite{Tadagi}.
Our results therefore also offers an understanding of this from the 
Euclidean approach. For the pure black hole and de Sitter configurations, 
our proposal is equivalent to the on-shell requirements but
it is interesting to note that the proposal also serves to give meaning 
to the concept of ``on-shell" for the Schwarzschild-de Sitter manifold;
and may be generalizable to even more complicated scenarios.

The Schwarzschild-de Sitter example also illustrates the much richer 
interplay among horizons, QFT and topology that can occur when the 
cosmological constant is {\it not neglected} in black hole scenarios.
It will be interesting to investigate the stability when back reactions
are taken into account. 
For instance since $\beta$ is given by the lowest common multiple
condition, it can vary wildly with deformations of $k_\pm$ if there
are no further restrictions. 
However, it is important to note that conservation of topological
Euler number implies that $(n_+ + n_-)$ should be constant. 
Thus within each topological sector where this number is conserved, 
$\beta$ varies inversely with $(k_- + k_+)$ (see Eq. (50)).
More chaotic behaviour can of course happen in quantum 
gravity when tunneling between different sectors and also 
violations of toplogical conservation laws are allowed. 

Finally, on possible interesting oscillating behaviour for 
evaporating black holes with cosmological constant, we feel that it is 
important to distinguish between the eternal and the evaporating case. 
The S-dS solution that we have is already a {\it four}-manifold. Its mass
parameter m does not increase or decrease with the time variable t, 
barring for instance superspace descriptions in quantum gravity where 
some other degree of freedom is chosen as ``time". 
An evaporating or anti-evaporating black hole with 
cosmological constant for which the size of the 
inner horizon increases or decreases with time is a different
four-manifold from the eternal case we have considered. Therefore the 
requirements on the periodicity (if there are any obvious ones following 
our prescription) are quite clearly different since the Kruskal 
extension will be that of another four-manifold. 
Thus our arguments do not necessarily 
imply that an evaporating black hole has arbitrarily large jumps in its 
Euclidean period. As for the superspace or quantum gravity context, 
we are neither able to prove nor disprove possible large jumps in the period.

\vskip 0.2in 						 
\section*{Acknowledgments} 
The research for this work has been supported in part 
by the Physics Department of Virginia Tech and 
the Natural Sciences and Engineering Research Council of Canada. 
We are grateful to L. N. Chang for discussions during the
progress of this work. 
C.S. would like to thank G. Kunstatter for helpful comments.

\end{document}